\documentclass[twocolumn,tightenlines,prd,floatfix,showpacs,superscriptaddress]{revtex4}
\usepackage{epsfig}
\usepackage{amssymb}
\usepackage{amsmath}
\usepackage{amsfonts}

\begin{document}

\title{Muon Fluxes and Showers 
from Dark Matter Annihilation in the Galactic Center}

\author{Arif Emre Erkoca}
\affiliation{Department of Physics, University of Arizona, Tucson, AZ 85721}

\author{Graciela Gelmini}
\affiliation{Department of Physics and Astronomy, University of California, 
Los Angeles, CA 90095} 

\author{Mary Hall Reno}
\affiliation{Department of Physics 
and Astronomy, University of Iowa, Iowa City, IA 52242}

\author{Ina Sarcevic}
\affiliation{Department of Physics, University of Arizona, Tucson, AZ 85721}
\affiliation{Department of Astronomy and Steward Observatory,
University of Arizona, Tucson, AZ 85721}

\begin{abstract}
We calculate contained and upward muon flux and contained 
shower event rates from neutrino interactions, when 
neutrinos are produced from  
 annihilation of the dark matter in the
Galactic Center.
We consider model-independent
direct neutrino production and secondary neutrino production from the
decay of taus, W bosons and bottom quarks produced in the annihilation of 
dark matter.  
We illustrate how muon flux from dark matter annihilation
has a very different shape than the muon flux from atmospheric neutrinos.  
We also discuss the dependence of 
the muon fluxes on the dark matter density profile and on 
the dark matter mass and of the total muon rates on the detector threshold.
We consider both the upward muon flux, when muons are created in the
rock below the detector, and the contained flux when muons are
created in the (ice) detector.  We also calculate the event rates for 
showers from neutrino interactions in the detector and show that 
the signal dominates over the background for $150 {\rm GeV} <m_\chi < 1$ TeV 
for $E_{sh}^{th} = 100$ GeV.

\end{abstract}


\maketitle

\section{Introduction}
Dark matter's presence is inferred from gravitational effects on
visible matter  at astronomical scales.
A wide range of  observational data  show that the dark matter is cold  or warm
(i.e. it became non-relativistic  before or at the time of galaxy formation)
 and composes about 23$\%$ of the
total density of the Universe \cite{DMobservations}.
 There are no viable candidates for dark matter within the standard model of elementary particles,
 but many in proposed extensions of the standard model.
Among these,
weakly interacting massive particles (WIMPs) of mass in the 100 GeV to several TeV range provide a natural explanation for the
observed  dark matter density  \cite{DMcandidates}. We are going to concentrate on WIMPs in this paper.

Although the detection  of dark matter particles may be possible at the Large Hadron Collider (LHC),
finding them in direct or indirect dark matter searches will be necessary
 to determine  if they are indeed stable on cosmological timescales and  how abundant they are at present  \cite{WIMP}.
 Many direct or indirect dark matter searches are being carried on  at present \cite{gelmini}.
Indirect dark matter searches look for WIMP annihilation (or sometimes decay) products, either photons  \cite{FGSTetc, INTEGRAL,WMAP-HAZE} or
anomalous cosmic rays, such as positrons and antiprotons \cite{AMS,HEAT,PAMELA, ATIC,PPB-BETS,HESS,FERMI}, or neutrinos   \cite{AMANDA,IceCube,KM3NeT}.
  For some years, observations of an excess in the positron fraction $e^+/(e^++e^-)$  by  HEAT (the High Energy Antimatter Telescope)
  \cite{HEAT}, a bright 511 keV gamma-ray line from the Galactic
Center  by INTEGRAL (the International Gamma Ray Astrophysics Laboratory) \cite{INTEGRAL} and
a possible unaccounted-for component of the foreground of WMAP around the galactic center,
the ``WMAP Haze"~\cite{WMAP-HAZE}  (among others)  have been considered possible hints of WIMP dark matter annihilations.
 
More recently, the  PAMELA satellite (Payload of Antimatter Matter Exploration and Light-nuclei Astrophysics)
  reported an excess in the positron fraction   in the energy range of 10-100 GeV with respect to what is
expected from cosmic rays secondaries \cite{PAMELA},  which confirmed the HEAT excess.
 Also ATIC (the Advanced Thin Ionization Calorimeter) and   PPB-BETS (the
Polar Patrol Balloon and Balloon borne Electron Telescope
with Scintillating fibers) observed
a bump in the $e^+$+$e^-$ flux from 200 to 800 GeV \cite{ATIC,PPB-BETS}, but this was  not confirmed by the
air Cherenkov telescope HESS \cite{HESS} and by the
 Fermi Gamma Ray Telescope. Fermi found a slight excess
in the e$^+$+e$^-$ flux between 200 GeV and 1 TeV
\cite{FERMI}.

Indirect searches for dark matter annihilations via
neutrinos with experiments such as AMANDA
(Antarctic Muon And Neutrino Detector
Array)  \cite{AMANDA} and
IceCube \cite{IceCube} also constrain dark matter models.
The cubic kilometer size neutrino
telescope (KM3NeT),
planned to be built at the bottom of
the Mediterranean Sea
\cite{KM3NeT}, will provide additional constraints, with its different
view of the sky and in particular, the galactic center. Many theoretical studies have concentrated on  the
indirect dark matter detection via neutrino signals \cite{theory,hisano,boost,boost_factor,Freese}.

The positron excess observed by PAMELA
may be explained by the presence
of particular astrophysical sources (e.g.,  pulsars) \cite{astro},   or by the annihilation \cite{leptophilic, non-standard_models} or  decay \cite{decay} of
dark matter particles.  If the observed anomalies in the
PAMELA and FERMI data are due to  dark matter annihilation, a larger annihilation rate than expected for typical thermal relics must be assumed. This enhancement
may happen due to either large inhomogeneities in the dark matter
 distribution near Earth (subhaloes) and/or
a larger annihilation cross section of the dark matter particles. This last possibility may happen if the dark matter particles
are not thermal relics  \cite{gelmini, non-standard_models},
 in which case they can have larger annihilation cross sections in the early Universe, or due to an enhancement of the annihilation cross section
only at  very low velocities \cite{sommerfeld}, which would not affect their annihilation in the early Universe.
Whatever its origin may be, the needed
enhancement is  quantified by a ``boost factor," $B$,
 ranging from 10 to 10$^4$ \cite{DMcandidates,boost,boost_factor,Freese}. The typical  WIMP thermal relic
annihilation cross section  is $\langle\sigma v\rangle =3\times10^{-26}\ {\rm cm}^3
{\rm s}^{-1}$.

WIMP models explaining the PAMELA positron excess must be peculiar in other aspects as well.
To avoid overproduction of antiprotons,
the dark matter annihilation or decay must proceed
dominantly to leptons. Moreover, the absence of a sharp shoulder in the electron plus positron spectrum
(that had been observed by ATIC) in the Fermi data  corresponding to an energy close to the
 parent dark matter particle mass means that the direct production of  electrons
 must be suppressed with respect to the production of electrons (and positrons) as secondaries.
 Final states including  $\tau$'s or  $\mu$'s  of dark matter  not lighter than
  1 TeV fit the PAMELA, HESS and Fermi data best \cite{Meade:2009iu}.
  These leptophilic dark matter candidates \cite{leptophilic} would copiously produce
 neutrinos  \cite{hisano}  whose fluxes are constrained by the observations of Super Kamiokande (SK)  \cite{SK}
toward the direction of the Galactic Center. Neutrinos with
energies of the order of the dark matter mass, $E_\nu \leq m_\chi$,
would propagate without being deflected towards the Earth.
However, during their travel, vacuum oscillation effects  would mix the three flavors.
Some fraction of the arriving muon neutrinos would be converted
into muons via charged current
interactions in the Earth which can be
detected in Earth based neutrino telescopes.

Neutrino signals in  underground or underwater detectors
of dark matter annihilation in the Galactic Center are
the subject of this paper. We calculate the neutrino induced
upward and contained muon flux,
as well as  the neutrino induced muon and shower event rates
due to dark matter annihilation
in the Galactic Center.  We take into account the muon propagation
in the Earth when evaluating the upward muon flux \cite{erkoca}
and study the energy range of muons for which upward muon events dominate
 over the contained ones. We show that the shape of upward muon fluxes
 differs significantly from the shape of the neutrino spectra at production, due to the smearing
 produced by neutrino interactions and muon propagation. The muon
 propagation shifts the flux to lower energies, while the contained muon flux increases
 with muon energy due to the linear energy dependence
  of the neutrino charged-current interaction.
 We consider different WIMPs annihilation channels
that contribute to the neutrino signal, including direct annihilation
to neutrinos, to charged leptons and to quarks or gauge bosons.
We evaluate rates of contained events and upward events,
of relevance to IceCube and future neutrino detectors like KM3NeT.

In the next section, we evaluate expressions for muon flux from the
incident neutrino flux interacting with the medium.
In Section III we present our results for muon flux and muon
event rates from the annihilation of the
 dark matter in the Galactic
Center compared with the atmospheric background and
 evaluate rates for hadronic and electromagnetic
showers.  Finally, in Section IV we
summarize and discuss our results.

\section{Muon Flux}

The neutrino flux at the Earth due to the annihilation of dark matter particles with mass $m_\chi$ in the Galactic Center 
is given by
\begin{equation}
\frac{d\phi_\nu}{dE_\nu}=R \times\left(\sum_F B_F\frac{dN_\nu^{F}}{dE_\nu}\right)
\end{equation}
where $R$ is the annihilation rate given by: 
\begin{equation}
R  = B\frac{\langle\sigma v\rangle}{8\pi m^2_\chi}\int 
d\Omega\int_{l.o.s}dl(\theta)\rho^2(l), 
\end{equation}
$dN_{\nu}^F/dE_\nu$ is the neutrino spectrum at the production for a given 
annihilation channel $F$
with branching fraction $B_F$, 
$B$ is the boost factor, 
$\rho(l)$ is 
 the dark matter density, integral is 
 over the 
line of sight (l.o.s) within a solid angle $\Delta\Omega$, 
centered in the Galactic Center.  The neutrino energy distribution,
$dN_\nu/dE_\nu$, depends on the particle produced. Some examples appear in 
Appendix A.
 For all of the 
evaluations
below, we take the dark matter annihilation cross section to have the typical thermal relic value
$\langle \sigma v\rangle=3\times10^{-26}\ {\rm cm}^3
{\rm s}^{-1}$.

For practical reasons the dimensionless quantity $\langle J_2\rangle_\Omega$ 
is defined in which the dark matter density profile $\rho(l)$ is embedded 
\cite{hisano},  
\begin{equation}
\langle J_2 \rangle_\Omega=\int \frac{d\Omega}{\Delta\Omega}\int_{l.o.s}\frac{dl(\theta)}{R_o}\left(\frac{\rho(l)}{\rho_o}\right)^2
\end{equation}
where $l(\theta)$ is the distance from us in the direction of
$\theta$ which is the cone half angle from the Galactic center, $R_o$ is the distance of the solar system from the Galactic Center  
and $\rho_o$ is the 
local density near the solar system, 
which are taken to be $R_o=8.5$ kpc and $\rho_o=0.3$ GeVcm$^{-3}$. As a practical matter, we consider two profiles, the Navarro-Frenk-White (NFW)\cite{NFW} profile and a 
cored isothermal profile. 
Some typical values for 
$\langle J_2\rangle_\Omega\Delta\Omega$ can be found in  Ref. \cite{hisano2}, where
$\langle J_2\rangle_\Omega\Delta\Omega$= 6.0(10.0) for $\theta=5^\circ(10^\circ)$ for the NFW profile, and 
$\langle J_2\rangle_\Omega\Delta\Omega$= 1.3(4.3) with $\theta=5^\circ(10^\circ)$ for 
the isothermal profile.

The high energy neutrinos coming from the 
Galactic Center then interact with the matter in the 
Earth and produce muons that traverse to the detector (upward events), or 
they interact in the detector producing muons or showers (contained events).  
Muon range or stopping distance, 
$R_\mu(E_\mu^i,E_{th})$, is given by 
\begin{equation}\label{range}
R_\mu(E_\mu^i,E_{th})=\frac{1}{\beta\rho}\log\left(\frac{\alpha+\beta{E_\mu^i}}{\alpha+\beta{E_{th}}}\right)
\end{equation}
where $\alpha$ corresponds to the ionization energy
loss and $\beta$ accounts for the
bremsstrahlung, pair production and photonuclear interactions. 
For example, for  a muon with initial energy 
$E_\mu^i \sim$ 1 TeV, 
when $E_{th}= 1\ {\rm GeV}$ the
muon range is 
 roughly $1$ km whereas the 
decay length 
of a muon with the same initial energy is much larger ($\sim$ a few thousand kilometers). For detectors
with a characteristic size of 1 km$^3$, contained events are most important for WIMP masses below about
1 TeV, while for smaller detectors like SuperK, upward events are relatively more important.

Using 
Eq.(1) and 
following the theoretical framework presented in Ref. \cite{erkoca}, 
the upward muon flux at the detector is given by 
\begin{eqnarray}\label{galactic_flux}
\frac{d\phi_\mu}{dE_\mu}&=&\int^{R_\mu(E_\mu^i,E_\mu)}_0 dz \int^{m_\chi}_{E^i_\mu}dE_\nu
\left(\frac{d\phi_\nu}{dE_\nu}\right)\\ \nonumber
&\times &
P_{surv}(E^i_\mu,E_\mu)\frac{dP_{CC}}{dzdE^i_\mu}\frac{dE^i_\mu}{dE_\mu}\\ \nonumber
&+& (\nu\rightarrow \bar{\nu}).
\end{eqnarray}
Here $P_{surv}$ accounts for muon energy loss in transit from its production position to the muon's entry into the detector. For an energy
independent energy loss parameter $\beta$, 
 the survival probability is 
\begin{equation}
P_{surv}(E_\mu^i,E_\mu) \simeq \Biggl( \frac{E_\mu}{E_\mu^i}\Biggr)^{\Gamma}
\Biggl( \frac{\alpha+\beta E_\mu^i}{\alpha+\beta E_\mu}\Biggr) ^\Gamma
\end{equation}
where $\Gamma = m_\mu/(c\tau_\mu\alpha \rho)$ in terms of the muon mass, muon
lifetime and the density of the medium $\rho$ in g/cm$^3$.

For production in the detector, the contained muon flux is
\begin{eqnarray}\label{galactic_flux2}
\frac{d\phi_\mu}{dE^i_\mu}&=&\int^D_0 dz \int^{m_\chi}_{E^i_\mu}dE_\nu
\left(\frac{d\phi_\nu}{dE_\nu}\right)\frac{dP_{CC}}{dzdE^i_\mu}\\ \nonumber
&+& (\nu\rightarrow \bar{\nu}).
\end{eqnarray}
where 
$D$ is the size of detector. The quantity 
$dP_{CC}$ 
is the probability for
a neutrino with energy 
$E_{\nu}$ to convert 
into a muon within the energy interval of 
$dE^i_\mu$ and over a distance $dz$:
\begin{equation}
\label{conversion}
dP_{CC} = dz\, dE^i_\mu \frac{N_A\rho}{2}\left(\frac{d\sigma^p_\nu(E_\nu,E^i_\mu)}
{dE^i_\mu}+(p\rightarrow{n})\right)\ , 
\end{equation}
where $N_A =6.022\times10^{23}$ is  Avogadro's number.  The differential cross sections
$d\sigma_{\nu}^{p,n}/dE^i_\mu$ are the weak scattering cross sections 
of (anti-)neutrinos on the nucleons,
which can be approximated by \cite{vissani}
\begin{equation}\label{weak_scattering}
\frac{d\sigma^{p,n}_{\nu,\overline{\nu}}}{dE^i_\mu}=
\frac{2m_pG^2_F}{\pi}\left(a^{p,n}_{\nu,\overline{\nu}}+
b^{p,n}_{\nu,\overline{\nu}}\left(\frac{E^i_\mu}{E_{\nu,\overline{\nu}}}
\right)^2\right)
\end{equation}
the parameters $a$ and $b$ for charged current
scattering are shown in Table \ref{table:sigmacc}.  

\begin{table}[t]
\begin{tabular}{|c|c|c | c|}
\hline
\hline
$a^p_\nu$ & 0.15 &$ b^p_\nu$ &  0.04\\
$a^p_{\bar{\nu}}$ & 0.04 & $b^p_{\bar{\nu}}$& 0.15 \\  
$a^n_\nu$ & 0.25 & $b^n_\nu $&  0.06\\
$a^n_{\bar{\nu}}$ & 0.06 & $b^n_{\bar{\nu}}$& 0.25 \\  
    \hline
         \hline
         \end{tabular}
\caption{Parameters for the charged current neutrino-nucleon differential cross section, as noted in
Ref. \cite{vissani}.}
\label{table:sigmacc}
\end{table}

\begin{table}[t]
\begin{tabular}{|c|c|c | c|}
\hline
\hline
$a^p_\nu$ & 0.058 &$ b^p_\nu$ &  0.022\\
$a^p_{\bar{\nu}}$ & 0.019 & $b^p_{\bar{\nu}}$& 0.064 \\  
$a^n_\nu$ & 0.064 & $b^n_\nu $&  0.019\\
$a^n_{\bar{\nu}}$ & 0.022 & $b^n_{\bar{\nu}}$& 0.058 \\  
    \hline
         \hline
         \end{tabular}
\caption{Parameters for the neutral current neutrino-nucleon differential cross section, as noted in
Ref. \cite{vissani}.}
\label{table:sigmanc}
\end{table}

Muon rates, 
 $N_\mu(m_\chi)$, are obtained by integrating Eqs.(5) and 
(7) over the muon energies, i.e.,
\begin{equation}
N_\mu(m_\chi)=\int^{m_\chi}_{E_{th}}\frac{d\phi_\mu}{dE_\mu}dE_\mu
\end{equation}
where $E_{th}$ is the muon detector threshold.

Another set of possible signals 
of dark matter are the showers produced in 
neutrino charged-current and neutral-current interaction in the 
detector.  
The contained shower flux in CC and NC interactions is given by 
 \cite{Dutta}: 
\begin{eqnarray}\label{showers}
\frac{d\phi}{dE_{sh}}&=&\int^D_0 dz \int^{m_\chi}_{E_{sh}}dE_\nu
\left(\frac{d\phi_\nu}{dE_\nu}\right)\frac{dP_{CC(NC)}}{dzdE_{sh}}\\ \nonumber
&+& (\nu\rightarrow \bar{\nu}).
\end{eqnarray}
where the shower energy is 
\begin{eqnarray}
E_{sh} \approx E_\nu-E_{\mu,\tau,e} 
\end{eqnarray}
The neutral current cross section can also be approximated with 
Eq. (\ref{weak_scattering}) where the parameters $a$ and $b$ appear 
in Table \ref{table:sigmanc}.

In the limit of the survival probability $P_{surv}$ going to unity,
the energy dependent flux can be calculated analytically when Eq. (9) is used for the neutrino-nucleon cross section. The analytic results for a variety of decay channels are shown in Appendix B.

\section{Results}

The direct production channel, $\chi\chi\rightarrow\nu_\mu\overline\nu_\mu$, 
 where $\chi$ is the WIMP, is the most promising channel for the
detection of dark matter annihilation, assuming an adequate annihilation
cross section, because of the monoenergetic neutrinos. 
A typical example of  a dark matter particle candidate which annihilates into a  neutrino pair
 is the lightest Kaluza-Klein particle. However, some particle candidates, for example  neutralinos and leptophilic dark matter, 
produce neutrinos only as secondary particles,
via the decay of  the particles into which the dark matter particles annihilate, such
as
$\mu^+\mu^-$, $\tau^+\tau^-$, $b \bar{b}$, $W^+ W^-$, etc.

In the first two figures, we present our results for the differential upward muon flux due to the annihilation of a dark matter particle via the direct production 
($\chi\chi\rightarrow\nu_\mu\overline\nu_\mu$) channel. 
To illustrate various contributions, we choose the dark matter particle
mass $m_{\chi} = $500 GeV, and for Fig. 1, the NFW dark matter density profile \cite{NFW}  and the boost factor $B=$200 
which is in the range of the boost factor values 
that explain the PAMELA data \cite{boost_factor}.
For Fig. 2, the dark matter density profile is the cored
isothermal profile and we use a boost factor $B=$ 800 to match the normalization of the NFW density profile for the  5$^\circ$ cone half angle.

We  show our 
results for two different choices of the
 cone half angle ($5^\circ$ and $10^\circ$) and 
compare them with the angle-averaged background due to the atmospheric 
neutrinos (in units of GeV$^{-1}$km$^{-2}$yr$^{-1}$sr$^{-1}$) 
\begin{eqnarray}\label{background}
\left(\frac{d\phi_\nu}{dE_\nu d\Omega}\right)_{ATM,avg} &=& N_0{E_\nu}^{-\gamma-1}(\frac{a}{bE_\nu}\ln(1+bE_\nu)+\nonumber\\
 &+ &\frac{c}{eE_\nu}\ln(1+eE_\nu)). 
\end{eqnarray}
which was obtained using the 
angle-dependent atmospheric neutrino flux parametrization in Ref. \cite{neutrino_background1}, 
\begin{eqnarray}
\frac{d\phi_\nu}{dE_\nu d\Omega} &=&
N_0{E_\nu}^{-\gamma-1}\nonumber\\
&\times &\left(\frac{a}{1+bE_\nu{cos\theta}}+\frac{c}{1+eE_\nu{cos\theta}}\right)\ .  
\end{eqnarray}
The values of the parameters 
$N_0$, $\gamma$, $a$, $b$, $c$ and $e$, given in Table III, 
were 
determined 
 by fitting angle-dependent atmospheric neutrino data from Ref. 
\cite{neutrino_background2}.  
The resulting final muon flux 
 with 
this approximated neutrino background is 
about $50\%$ larger (smaller) than that from the 
vertical (horizontal) atmospheric neutrinos.      

For a $10^\circ$ cone half angle, the signal 
dominates over the background in the 
range 180 GeV$<E_\mu<$420 GeV for the NFW profile. We note that 
the background signal is suppressed 
more than the dark matter signal with the 
decrease in the cone of half angle. As a comparison, for a $5^\circ$ 
 cone half angle the signal exceeds the background in a wider range of energies, 60 GeV$<E_\mu<$480 GeV.

\begin{table}[t]
\begin{tabular}{|c|c|}
\hline
\hline
$\gamma $& 1.74\\ 
$a$ & 0.018\\
$b$ & 0.024\ GeV$^{-1}$\\  
$c$ & 0.0069\\
$e$ & 0.00139 GeV$^{-1}$\\
$ N_0$ & $
         \begin{array}{lr}
         1.95\times10^{17}&  \mbox{for}\;\;\nu \\
         1.35\times10^{17}&  \;\mbox{for}\;\;\overline\nu.
         \end{array}$
         \\
         \hline
         \hline
         \end{tabular}
\caption{Parameters for the atmospheric $\nu_\mu$ and $\bar{\nu}_\mu$ flux, 
in units of
GeV$^{-1}{\rm km}^{-2}{\rm yr}^{-1}{\rm sr}^{-1}$.}
\label{table:atm}
\end{table}

\begin{figure}[h]
\begin{center}
\epsfig{file=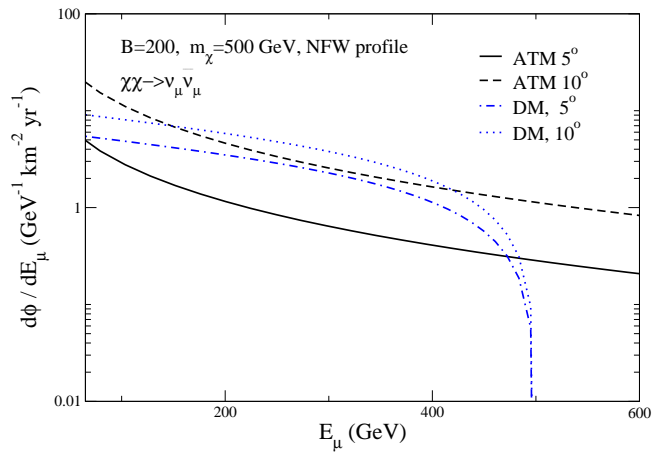,width=2.9in,angle=270}
\end{center}
\caption{Upward muon flux obtained from dark matter annihilation into neutrinos in the Galactic Center, for a cone half angle ($\theta$) of 
5$^\circ$ 
(dot-dashed) and 10$^\circ$ (dotted). The background upward muon fluxes due to (angle-averaged) atmospheric neutrinos are shown with the solid (for $\theta=5^\circ$) 
and the dashed (for $\theta=10^\circ$) curves. The NFW dark matter profile
is used, along with a boost factor $B=200$ and $m_\chi=500$ GeV.}
\label{}
\end{figure}

From Fig. 2, we note that in case of 
the isothermal profile for the dark matter in which there
is
a relatively less dense core region 
in the isothermal profile,  by increasing the cone half 
angle from 5$^\circ$ to 10$^\circ$, there is
an almost equal enhancement of the upward muon fluxes 
from the atmospheric neutrino background and from the dark matter annihilation in the center of 
the galaxy. 
For the set of the parameters that we choose here, the dark matter signal becomes larger than the background in the energy ranges of 100 
GeV$<E_\mu<$470 GeV and 70 GeV$<E_\mu<$480 GeV for the cone half angles  $10^\circ$ and $5^\circ$, respectively.

\begin{figure}[h]
\begin{center}
\epsfig{file=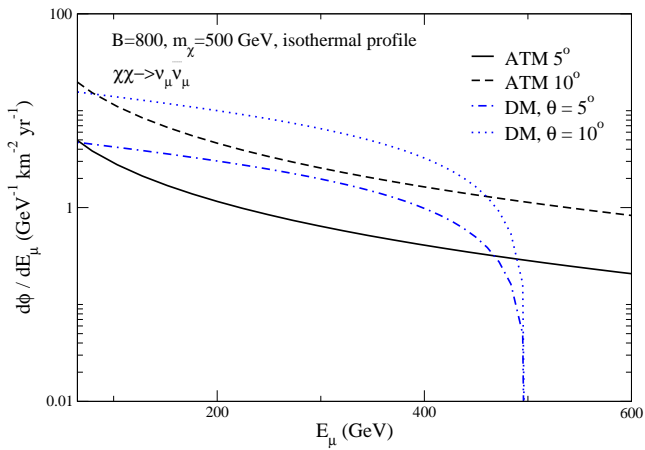,width=2.9in,angle=270}
\end{center}
\caption{Same as Fig. 1 but for the cored isothermal dark matter density profile and a boost factor $B=800$.}
\label{}
\end{figure}

Fig. 3 shows the dependence of the 
differential muon fluxes from dark matter annihilation via the direct production channel for $m_\chi=200,\ 500$ and 800 GeV.
We again consider the NFW profile, a fixed boost factor ($B=200$) and a fixed cone  half angle ($\theta=5^\circ$). 
The figure shows the upward flux as well as the contained flux 
assuming a detector size $D=1$ km in Eq. (7). 
We find that regardless of the mass dependence,the upward event spectrum is a decreasing function of the muon energy whereas the corresponding 
spectrum of the contained events increases with the muon energy up to the cut-off set by the initial neutrino energy. In 
our calculations, we assume that the dark matter particles annihilate at rest and thus the neutrino energy for this decay mode can be set to the 
rest mass of the dark matter particle,  $E_\mu = m_\chi$. 

The signal for the muon flux 
from the contained events has a stronger suppression with the increase in the 
dark matter mass than for the upward muon events.
This is due to the $m^{-2}_\chi$ dependence in Eq. (2). 
The mass dependence for upward events is more complex because
of the mass dependence in the upper limit of the $z$ integration in 
Eq. (5). A large mass $m_\chi$ (and therefore higher $E_\nu$) produces a higher
energy muon which has a longer range in the rock below the detector. For 
example, for $E_\mu>380$ GeV, the upward event signal from the 
annihilation of the dark matter particle with mass $m_\chi=800$ GeV dominates over the 
one from that of the dark matter particle with mass $m_\chi=500$ GeV.  

For a wide 
range of muon energies, 
the dark matter signal is above the atmospheric 
background both for contained and upward events in the
$\chi\chi\rightarrow\nu_\mu\overline\nu_\mu$ channel with the boost factor
used here. 
We find that for a given dark matter mass 
the contained events exceed the upward ones in the range $E_\mu\ge 0.6 m_\chi$.

\begin{figure}[h]
\begin{center}
\epsfig{file=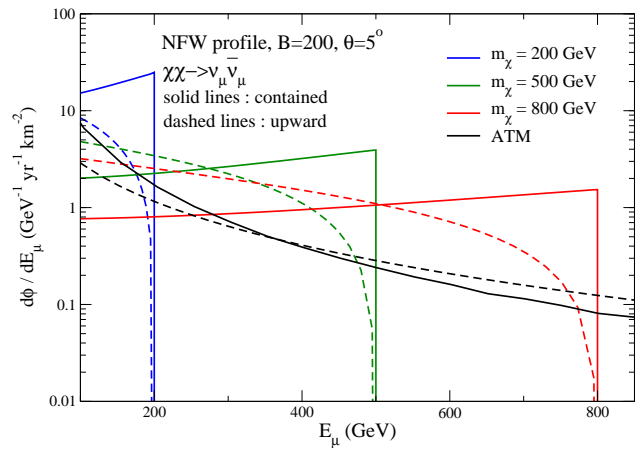,width=2.9in,angle=270}
\end{center}  
\caption{Muon flux due to the dark matter annihilation into neutrinos in the 
Galactic Center for different dark matter masses,
curves correspond to the dark matter masses of 200 GeV, 500 GeV and 800 GeV, respectively. The corresponding backgrounds are also shown. All the solid lines correspond to 
the contained events with $D=1$ km,
whereas the dashed ones to upward events.}
\label{}
\end{figure}

In Fig. 4, we present our results for the differential muon flux due to 
$\chi\chi\rightarrow\tau^+\tau^-$ 
channel. This channel is characteristic
of all three-body decays into neutrinos (secondary neutrinos). 
Again shown are the upward
and contained signals from $m_\chi=200,\ 500$ and 800 GeV with the
NFW profile and $B=200$.

Note that
in the case of  secondary neutrinos, the signal for both upward and contained
events decrease as the muon energy increases,
and for a fixed $m_\chi$, the contained events, in general, dominate over the upward
 events for muon energies $100 {\rm GeV} \le E_\mu \le m_\chi$. This is a consequence of considering
a detector size of $D=1$ km, a size larger than the range of a muon
with an energy of less than 1 TeV.  
The figure shows that even for a half angle of $5^\circ$, in case of
NFW profile one would need a boost factor on the order of
about $2000$ for the
 dark matter
signals from the secondary neutrinos to be above
 the atmospheric background.

\begin{figure}[h]
\begin{center}
\epsfig{file=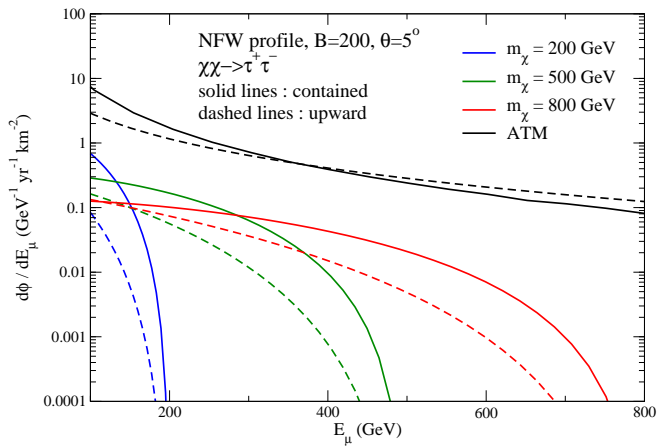,width=2.9in,angle=270}
\end{center}
\caption{Muon fluxes due to the secondary neutrinos produced through the dark matter annihilation into tau particles in the Galactic center for 
different dark matter
masses; $m_\chi=200, 500$ and 800 GeV. The
solid (dashed) curves correspond to contained (upward) events.}
\label{}  
\end{figure}

Measurement of the
muon flux can also be used to distinguish different dark matter models, as
seen in Fig. 5 where we
compare signals from
different annihilation channels: $\chi\chi\rightarrow{W}^+{W}^-$, 
$\chi\chi\rightarrow{\tau}^+{\tau}^-$ and 
$\chi\chi\rightarrow{b}\overline{b}$ for the NFW profile, with $B=200$,
the half angle equal $5^\circ$ and $m_\chi=500$ GeV.
The signals from 
the b-quark and the tau decay modes differ only by an overall 
factor which is close to the ratio of the decay branching 
fractions of the corresponding modes given in the
Appendix I.  However, for the $W$ decay, being a 2-body decay, 
the shape of the differential muon spectrum is quite different than those 
of the b-quark and tau which are 
both 3-body decay modes. 
This indicates that muon flux from the 
secondary neutrinos as a by-product of the 
dark matter annihilation can also be useful in 
discriminating different dark matter models.      

\begin{figure}[h]
\begin{center}
\epsfig{file=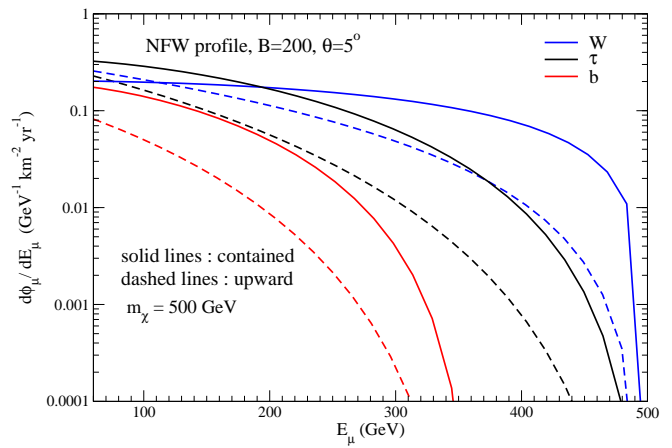,width=2.9in,angle=270}
\end{center}
\caption{Muon fluxes due to the secondary neutrinos produced through the dark matter annihilation into W bosons, tau particles,
and bottom quarks in the Galactic Center. The solid (dashed) lines for each channel correspond to contained (upward) events. The detector size is
taken to be $D=1$ km, and the cone half angle is $\theta=5^\circ$ for
the NFW profile with $B=200$.} 
\label{}
\end{figure}

We now turn to the total rate of upward and contained muons produced
by $\nu_\mu +\bar{\nu}_\mu$ from direct dark matter annihilation to 
neutrinos.
Integrating the differential fluxes over the final muon energy,
we obtain the muon rate from the annihilation of the dark matter as a
function of the
mass $m_\chi$ (Fig.6) for the NFW profile with 
$B=200$ and $\theta=5^\circ$. Here, the threshold energy is
taken to be $E_{th}=80$ GeV. Due to the 
finite size of the detector ($D=1$ km), and $m_\chi^{-2}$ dependence of 
the annihilation rate, the
signal for 
the contained events decreases with increasing the dark matter mass.  
On the other hand 
for upward events, 
heavier dark matter particles yield more energetic
neutrinos which makes a larger portion of 
muons in the rock below the detector to
contribute to the final muon flux.  This effect combined with the 
energy dependence of the neutrino charged-current cross section, results in 
increasing muon rate up to 
$m_\chi = 650$ GeV, at which point the $m_\chi^{-2}$ dependence of the 
annihilation rate takes over resulting in 
slow decrease of the muon rate.   
Comparison of contained and upward muon rates presented 
in Fig. 6 indicates that 
 for $m_\chi\le500$ GeV the signal 
from the contained events still dominates 
over the signal from the upward events.  
Even though 
 the signal depends weakly on the value of the 
threshold energy, 
 the background is very sensitive to it due to the large 
contribution from the low energy 
atmospheric neutrinos.  The signal to background ratio 
increases with increasing the muon energy threshold.  
We obtain the same results for the isothermal dark matter density 
halo profile if the boost factor is taken to be 800 for the same 
cone half angle of 5$^\circ$.  

\begin{figure}[h]
\begin{center}
\epsfig{file=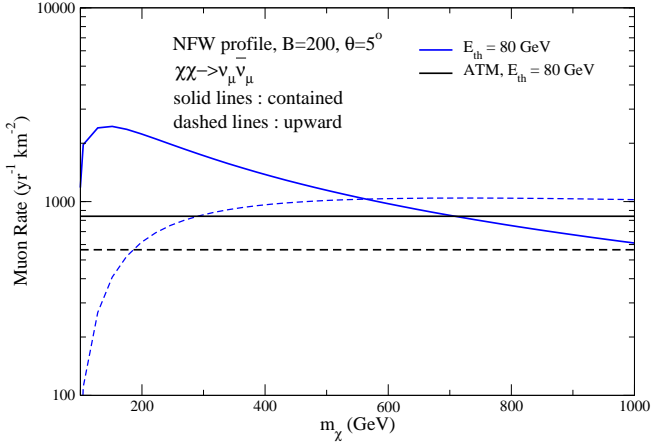,width=2.9in,angle=270}
\end{center}
\caption{
Total muon fluxes due to the dark matter annihilation into neutrinos in the Galactic Center. 
The solid (dashed) lines correspond to contained (upward) events.} 
\label{}
\end{figure}

\begin{figure}[h]
\begin{center}
\epsfig{file=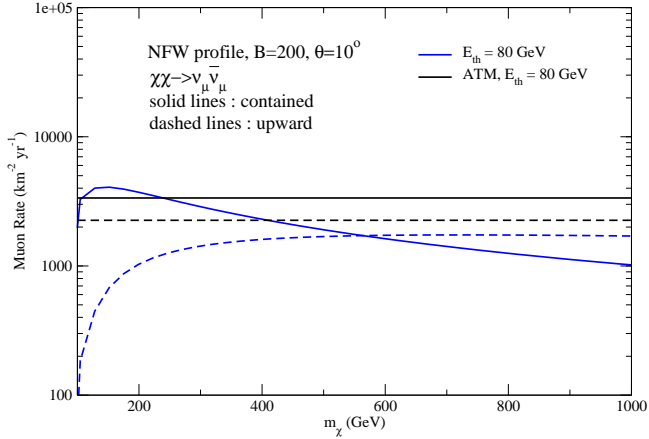,width=2.9in,angle=270}
\end{center}
\caption{Same as Fig.(6) but for 10$^\circ$ cone half angle.}
\label{}
\end{figure}

In Fig. 7 we show our results for 
the 10$^\circ$ cone half angle.  We note that in case of contained events the signal dominates  
over the background for $100$ GeV $\le m_\chi\le 200$ GeV, when 
the threshold energy is $80$ GeV. For upward events, signal is below the 
background for all $m_\chi$.  
The 
isothermal dark matter density halo profile gives 
larger signal than obtained with the NFW profile by about a factor of 
$2$, due to its larger increase of 
$\langle J_2\rangle_\Omega$ for 10$^\circ$ relative to 5$^\circ$.  

\begin{figure}[h]
\begin{center}
\epsfig{file=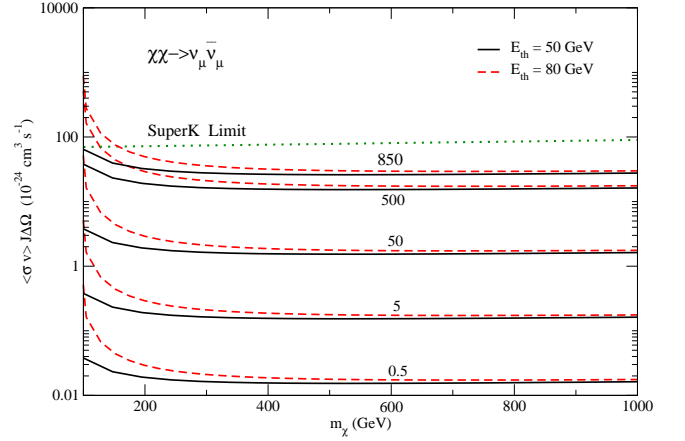,width=2.9in,angle=270}
\end{center}
\caption{Upward muon events curves, $N_{\mu}=(0.5,5,50,500,850)$km$^{-2}$yr$^{-1}$,
 for the energy
threshold of 50 GeV and 80 GeV are shown by the
solid and the dashed lines, respectively. The boost factor is set to  
be unity and the cone half angle is chosen to be 5$^\circ$.} 
\label{}
\end{figure}

In Fig. 8 we show contour plots
for upward muon events, 
$N_{\mu}=(0.5,5,50,500,850)$km$^{-2}$yr$^{-1}$.  
The solid (dashed)
lines
correspond to the muon energy threshold of
50 (80) GeV.  We also calculate that 
$N_\mu=714 (516)$km$^{-2}$yr$^{-1}$ for the upward muon events
 due to the atmospheric muon neutrinos
for the muon energy threshold of 50 (80) GeV. We find that for a fixed cone 
half angle the annihilation cross section does not depend on
$m_\chi$ for $m_\chi>200$ GeV to produce a given total muon
flux since the decrease in the annihilation rate with $m_\chi$ is
compensated with the increase in the muon range and neutrino cross section with $m_\chi$.
The
dependence on the choice of the threshold is also negligible. However,
for low mass
dark matter particles, higher values of the
annihilation cross sections are required
in order to have the same total muon flux. This is due to the fact that
the neutrinos originated from this low mass dark matter annihilation mostly
contribute to the muon flux at energies less than the thresholds we choose.
The parameter space above the dotted line is excluded at $90\%$ C.L. by
Super-Kamiokande observations toward the direction of
the Galactic Center with a cone half angle of $5^\circ$
\cite{SK}.

The dominant atmospheric neutrino flavor at neutrino energies above 
$40$ GeV is $\nu_\mu$ which produces track-like events through charged 
current interactions in the neutrino telescopes.  Identifying track-like 
events could reduce the background substantially.  Recently it has been 
argued that 
IceCube+DeepCore will be able to 
put constraints on dark matter properties 
in a more efficient way by just analyzing the cascade (i.e shower) events 
which are due to charged current interactions of $\nu_{e,\tau}$ and 
the neutral current interactions of 
the all neutrino flavors \cite{Freese}.  
Since the weak scattering cross sections are independent of the flavors, the 
signal-to-background ratio is enhanced in shower events since $\nu_\mu$ can 
only contribute to the shower events through neutral current interactions 
where the cross section about 1/3 of the 
charged current cross section. 

In Fig. 9, we show hadronic shower rates as 
a function of $m_\chi$ from
neutral-current and charged-current interactions of 
muon neutrinos and antineutrinos.
These rates are the same for any other neutrino flavor with
a democratic $\chi \chi \to \nu\bar{\nu}$ annihilation rate.
Also shown is the 
hadronic shower rate due to the 
atmospheric muon 
neutrinos;  $N^{atm}_{sh}=524 (168)$km$^{-2}$yr$^{-1}$ for the charged current (neutral current) interactions. The shower threshold is taken to be 100 GeV.
We note that the 
background due to the atmospheric electron and tau neutrinos 
is much smaller than for the muon neutrinos, so the signal to background 
would not change much here
when all the neutrino flavors were included.  

We also 
evaluate the
electromagnetic shower rate as a function of
$m_\chi$ due to electrons produced by the 
charged-current interactions of $\nu_e$, with an electromagnetic
shower threshold set at 100 GeV.  
The atmospheric shower rate is evaluated using
the
atmospheric $\nu_e$ and $\overline{\nu}_e$ flux for 
an effective zenith angle $0.4<\cos\theta_z<0.5$, which roughly 
corresponds to the angle describing the position 
of the Galactic Center relative to the IceCube, 
\begin{eqnarray}  
\left(\frac{d\phi}{dEd\Omega}\right)_{\nu_e} &=& 
\frac{500.0}{({\rm GeV m^2 s\, sr})} \Biggl(
\frac{E}{{\rm GeV}}\Biggr)^{-3.57}
 \nonumber\\ 
\left(\frac{d\phi}{dEd\Omega}\right)_{\overline{\nu}_e} &=& 
\frac{382.6}{(\rm{GeV m^2 s\, sr})} \Biggl(\frac{E}{{\rm GeV}}\Biggr)^
{-3.57}
. 
\end{eqnarray}  
From Fig. 10 we see that the 
signal-to-background ratio is increased 
for the 
electromagnetic showers relative to hadronic showers (see Fig. 9) 
mainly due to a very small atmospheric electron neutrino flux which 
is about $34 {\rm km}^{-2} {\rm yr}^{-1}$. For secondary 
electron neutrinos from the decay of taus 
which are produced via $\chi\chi\rightarrow\tau^+\tau^-$, 
the signal becomes comparable 
to the background.

For the future neutrino detector which is positioned in the 
northern hemisphere, such as KM3Net, the relevant background would be 
coming from almost horizontal showers, which is 
about a factor of 
three to four times larger than the flux given by 
Eq. (14), giving 
approximately electromagnetic shower flux of 
 100 ${\rm km}^{-2} {\rm yr}^{-1}$.  

\begin{figure}[h]
\begin{center}
\epsfig{file=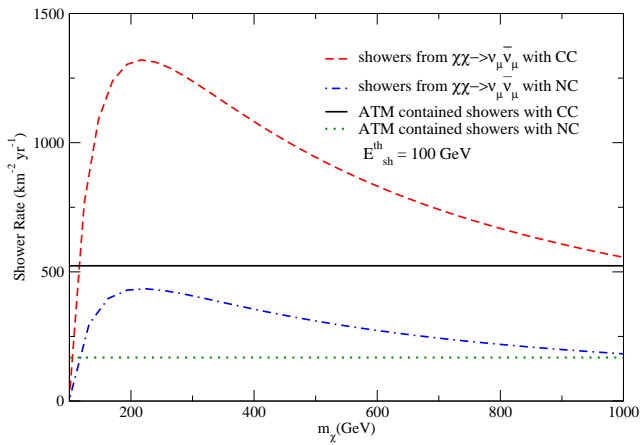,width=2.9in,angle=270}
\end{center}  
\caption{Hadronic shower rates for charged-current (dashed) and neutral current (dot-dashed) interactions of 
$\nu_\mu + 
\overline{\nu}_\mu$ when muon neutrinos are 
 produced directly from the dark matter 
annihilation in the Galactic Center, 
compared with the 
atmospheric background.
The NFW profile, with $B=200$, $\theta = 5^\circ$ and
$D=1$ km are used.
}
\label{}
\end{figure}

\begin{figure}[h]
\begin{center}
\epsfig{file=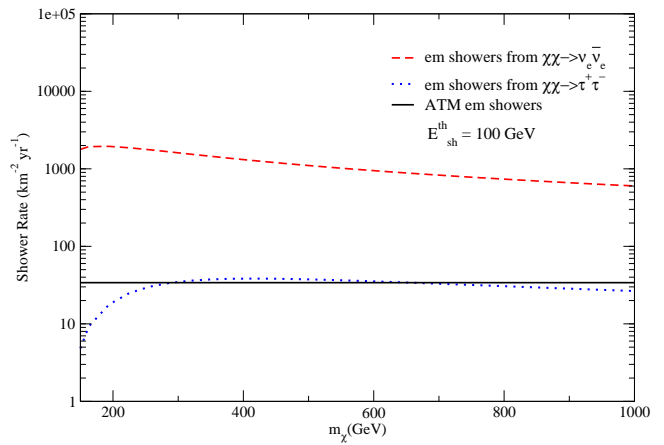,width=2.9in,angle=270}
\end{center}
\caption{Electromagnetic shower rates 
as a function of $m_\chi$ for $\nu_e + \overline{\nu}_e$ 
charged-current interactions when electron neutrinos are produced 
directly in the 
annihilation of dark matter in the Galactic Center, compared with the 
atmospheric background for shower energies above 100 GeV. 
The NFW profile, with $B=200$, $\theta = 5^\circ$ and $D=1$ km
are used.}
\label{}
\end{figure}

In Fig. 11 and 12, we present the contour plots for contained showers
with the energy threshold of 100 GeV. The main difference between the showers and the upward muons
appears for $m_\chi>200$ GeV where for a given total number of shower
events higher annihilation cross sections is required with the increase
in $m_\chi$. This is due to the contained event nature of the
shower events which are all produced inside the detector with finite size.
Thus, in contrast to the case for the upward muon events that we discussed earlier,
the strong suppression of the annihilation rate with $m_\chi$ can not
be compensated because of the finite size of the detector.
The charged current showers actually require a smaller annihilation 
cross sections
in order to produce the same number of total shower events that neutral current showers produce
for a fixed $m_\chi$ due to the larger weak scattering cross sections.

\begin{figure}[h]
\begin{center}
\epsfig{file=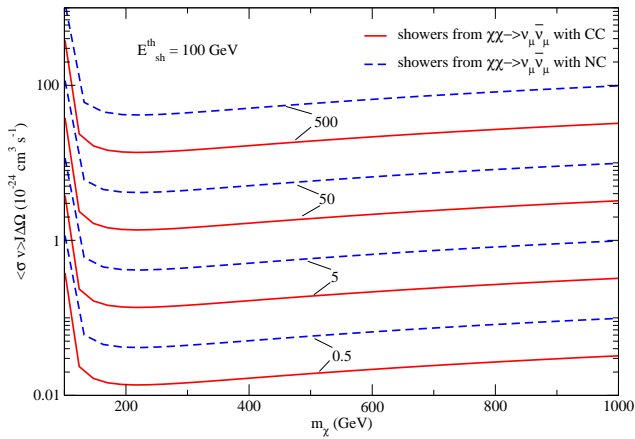,width=2.9in,angle=270}
\end{center}
\caption{Hadronic shower events curves, $N^h_{sh}=(0.5,5,50,500)$km$^{-2}$yr$^{-1}$,
 for the charged current (solid) and neutral current (dashed)
processes, for a NFW dark matter density profile, 
a 5$^\circ$ cone half angle, the boost factor set to be unity and $D=1$ km.}
\label{}
\end{figure}

\begin{figure}[h]
\begin{center}
\epsfig{file=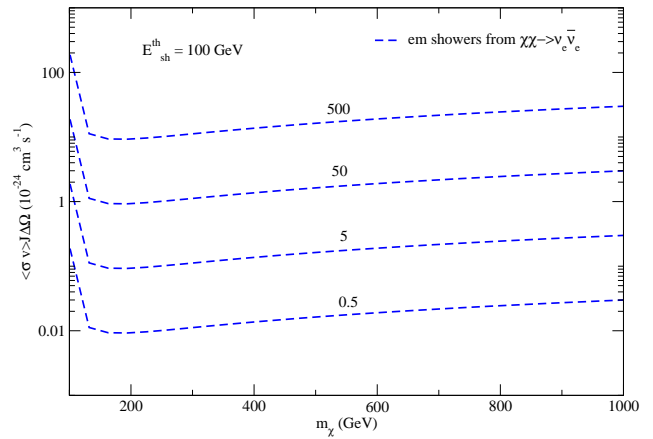,width=2.9in,angle=270}
\end{center}
\caption{Electromagnetic shower events curves, $N^{em}_{sh}=(0.5,5,50,500)$km$^{-2}$yr$^{-1}$
for a NFW dark matter density profile, a 5$^\circ$ cone half angle, the 
boost factor set to be unity and $D=1$ km.}

\label{}
\end{figure}  

\begin{figure}[h]
\begin{center}
\epsfig{file=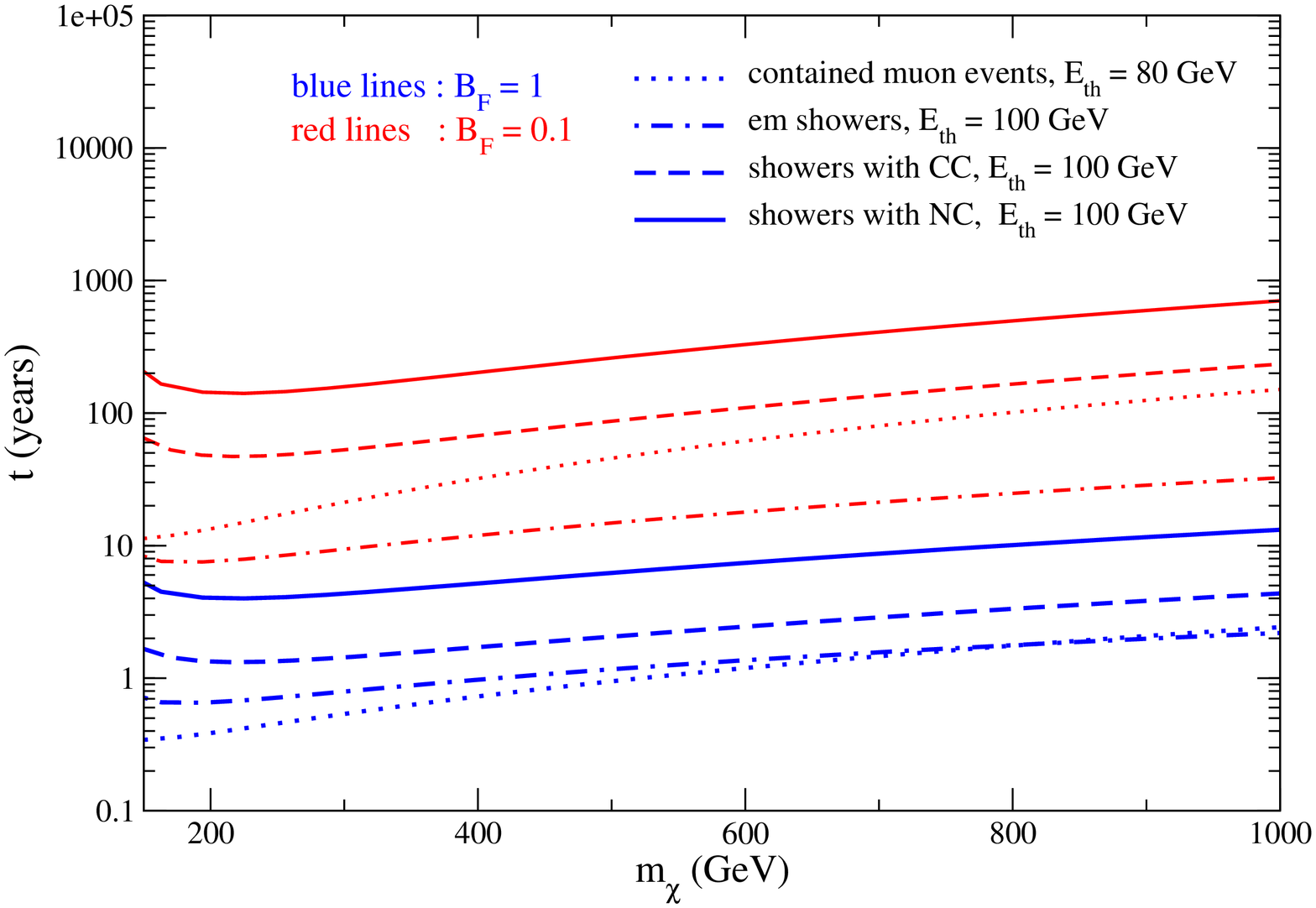,width=2.9in,angle=270}
\end{center}
\caption{Time as a function of dark matter mass, $m_\chi$, 
for the direct neutrino production channel ($\chi \chi \rightarrow 
\nu \bar{\nu}$) to reach
a 5$\sigma$ detection level for IceCube+DeepCore detector.  
The curves correspond to 
hadronic showers (solid for neutral current, dashed for charged current interactions), electromagnetic showers (dotted) and the 
contained muon events (dot-dashed). $B_F= 1 (0.1)$ for the lower (upper) 
 curves, 
 the boost factor is taken to be 200 and the cone half angle is 
 $5^\circ$ for all curves.}
\label{}
\end{figure}

The signal 
detection significance can be evaluated using 
\begin{equation}\label{significance}
S=\frac{N_s}{\sqrt{(N_s + N_b)}}, 
\end{equation}
where $N_s$ corresponds to the number of events for the 
signal, 
while $N_b$ is the background.  
We 
 obtain the time it would take to 
observe a 
5$\sigma$ effect using 
our results for the contained muon events (Fig. 6), hadronic showers (Fig. 9) and electromagnetic showers (Fig 10), 

\begin{equation}\label{significance}
t=\frac{25(N_s + N_b)}{N_s^2 V}
\end{equation}
where $V=0.04(0.02) {\rm km}^3$ is the effective volume of 
IceCube+DeepCore for the track-like (shower) events.  
  In Fig. 13, we 
show the observation time ($t$) required for IceCube+DeepCore detector to
detect or exclude the dark matter signal via 
the direct production channel at a 5$\sigma$ level.  
Here, we again use fixed boost factor ($B=200$)
and cone half angle ($\theta=5^\circ$). 
Our results, when we take $B_F=1$ for the direct production channel, suggest 
that in less than two years of observation
IceCube+DeepCore will be able to reach a 5$\sigma$ detection for the contained 
muon and 
electromagnetic shower events for a wide range of $m_\chi$. Decreasing the 
branching fraction by an order of magnitude
increases the observation time significantly in order to reach the 
same significance.  For instance, $t\simeq {\rm 10 - 50} $ years, 
for $150 {\rm GeV} \ge m_\chi \le 500 {\rm GeV}$ in the case of 
contained muon events, 
and somewhat shorter for the electromagnetic showers.

\begin{figure}[h]
\begin{center}
\epsfig{file=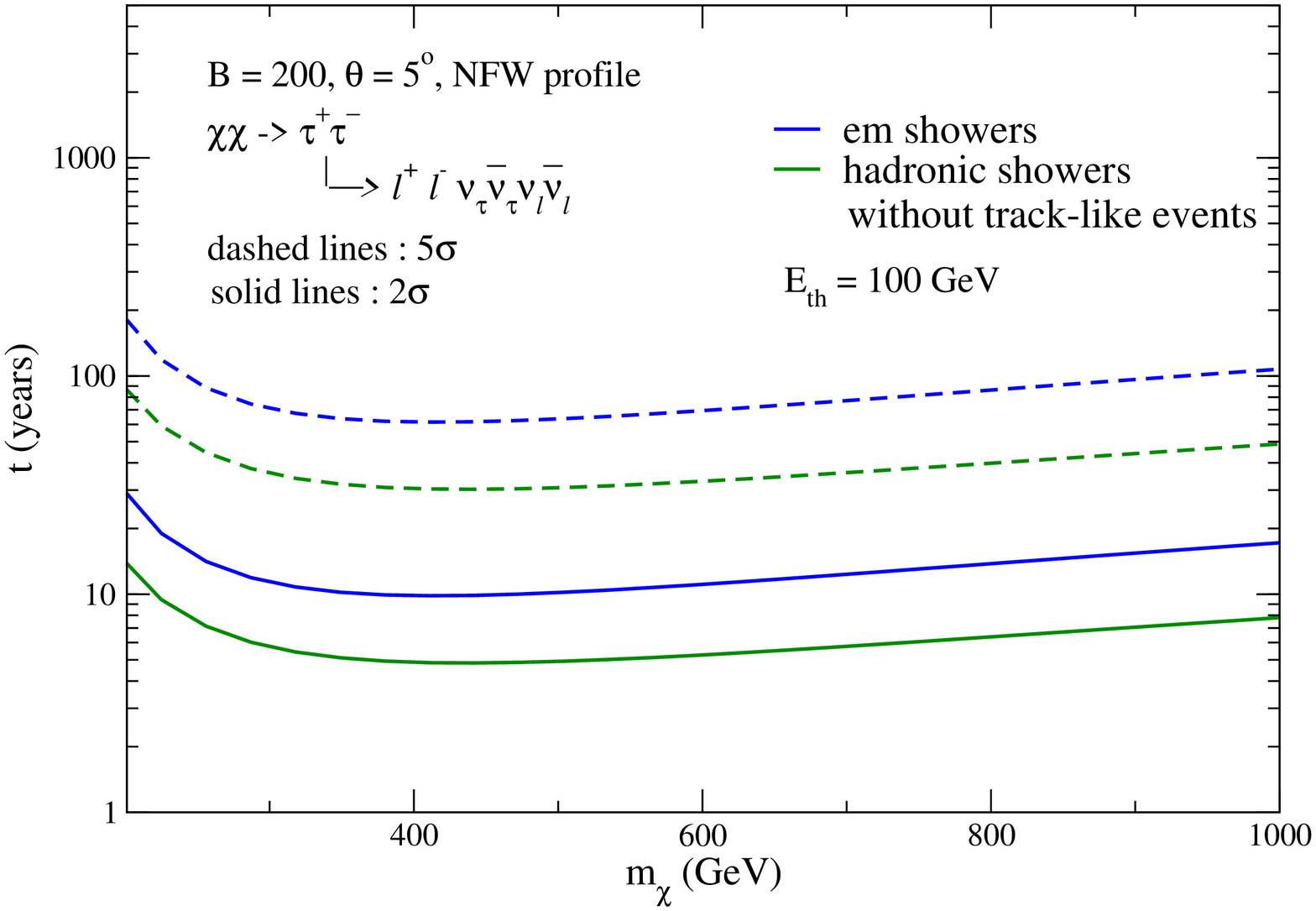,width=2.9in,angle=270}
\end{center}
\caption{Time as a function of dark matter mass, $m_\chi$, for the secondary neutrino production channel
$\chi \chi \rightarrow \tau^+ \tau^- \rightarrow
l^{+}  l^{-}  \nu_\tau \bar{\nu}_\tau \nu_l \bar{\nu}_l$ to reach
a 2$\sigma$ (solid curves) or a 5$\sigma$ (dashed curves) detection level
when measuring electromagnetic showers (top curves) and
hadronic showers without charged track-like events (lower curves).}
\label{}
\end{figure}

In the case of secondary neutrino production,
when neutrinos are produced from tau decays, and taus are
products of dark matter annihilation, 
 these neutrinos can interact
inside the detector producing hadronic and electromagnetic showers,
in addition to muon neutrinos producing muons via charge-current
interactions.  In Fig. 14 we show that IceCube+DeepCore detector
could potentially detect a 2$\sigma$ effect in 5 (8) years for
$m_\chi=300$ GeV ($1$TeV), in case
of excluding muon-like events.
To reach
a 2$\sigma$ detection for the electromagnetic showers due to the
secondary electron neutrinos
IceCube+DeepCore will need about $10-20$ years of observation for
 $250$ GeV $\le m_\chi\le 1$ TeV. When muon-like events are included, 
the observation times 
for the hadronic showers become similar to those for the 
electromagnetic showers.
The time needed for a 5$\sigma$
effect for hadronic (electromagnetic) showers
is almost an order of magnitude longer than for
a 2$\sigma$ effect.

Comparing the secondary and direct production (Fig. 13)
one sees that it takes longer (by about one order of magnitude) to detect
showers
from secondary neutrinos that to detect showers from primary neutrinos.
This is because of the different shape of the shower energy
distributions: for direct neutrinos it increases with energy and for
secondary neutrinos it decreases with energy.

Since the angular resolution for showers is expected to be much
worse than for muons, for the angular resolution of
 $30^\circ$, the
number of signal events will be larger by a factor of 6, while
the background will increase by a factor of 35, which results in
reducing the time it would take IceCube+DeepCore to see a 2$\sigma$
effect to 3 years for hadronic showers without 
track-like events. This is in qualitative agreement with
the results presented in Ref. \cite{freese2}.

For dark matter models in which neutrinos are decay products of
 taus produced in the dark matter annihilation, looking for
contained hadronic showers in IceCube+DeepCore seems promising
to detect a signal at the 2 sigma level, assuming the NFW dark matter halo
profile and a boost factor $B=200$.

In Table (\ref{table:summary}) we give a summary of our results for the event rates for various 
dark matter masses. We consider the direct production of neutrinos ($\chi\chi\rightarrow\nu\overline\nu$) 
and the neutrinos from the tau decay ($\chi \chi \rightarrow \tau^+ \tau^- \rightarrow
l^{+}  l^{-}  \nu_\tau \bar{\nu}_\tau \nu_l \bar{\nu}_l$). We classify the event rates as contained (ct) and upward (up)
for the track-like muon ($\mu$) events, and depending on the type of the interaction involved 
charged current (CC), neutral current (NC) and electromagnetic (em) for 
the shower events. Two different cone half angles are 
chosen, $\theta=5^\circ$ and $\theta=10^\circ$, and 
the threshold energy for the track-like muon (shower) 
events are set to be 80 (100) GeV.
We also show the atmospheric neutrino background for the track-like muon and for 
the shower events.

\begin{table}[h]
\centering
\begin{tabular}{l|ccccccccc}
\hline\hline
  &\multicolumn{9}{c}{$m_\chi$ (GeV)}\\[-1ex]
  & 200 & 300 & 400 & 500 & 600 & 700 & 800 & 900 & 1000 \\
$\chi\chi\rightarrow\nu\overline\nu$  & & & & & & & &  & \\
\hline
$N^\mu_{ct} (5^\circ)$  & 2240 & 1750 & 1385 & 1135 & 976 & 850 & 750 & 670 & 611  \\
$N^\mu_{ct} (10^\circ)$ & 3808 & 2975 & 2355 & 1930 & 1659 & 1445 & 1275 & 1139 & 1039  \\
$N^\mu_{up} (5^\circ)$ & 615 & 850 & 960 & 1010 & 1035 & 1042 & 1040 & 1033 & 1023  \\
$N^\mu_{up} (10^\circ)$ & 1046 & 1445 & 1632 & 1717 & 1760 & 1771 & 1768 & 1756 & 1739  \\
$N^{NC}_{sh} (5^\circ)$ & 430 & 400 & 355 & 310 & 274 & 240 & 220 & 200 & 182   \\
$N^{NC}_{sh} (10^\circ)$ & 731 & 680 & 604 & 527 & 466 & 408 & 374 & 340 & 309   \\
$N^{CC}_{sh} (5^\circ)$ & 1310 & 1230 & 1080 & 935 & 830 & 741 & 665 & 605 & 556   \\
$N^{CC}_{sh} (10^\circ)$ & 2227 & 2091 & 1836 & 1590 & 1411 & 1260 & 1131 & 1029 & 945   \\
$N^{em}_{sh} (5^\circ)$ & 1920 & 1600 & 1300 & 1100 & 950 & 820 & 730 & 660 & 600   \\
$N^{em}_{sh} (10^\circ)$ & 3264 & 2720 & 2210 & 1870 & 1615 & 1394 & 1241 & 1122 & 1020   \\
\hline
$\chi\chi\rightarrow\tau^+\tau^-$  & & & & & & & & & \\
$N^{NC}_{sh} (5^\circ)$ & 17 & 28 & 33 & 33 & 32 & 31 & 28 & 27 & 24  \\
$N^{NC}_{sh} (10^\circ)$ & 29 & 48 & 56 & 56 & 54 & 53 & 48 & 46 & 41  \\
$N^{CC}_{sh} (5^\circ)$ & 39 & 66 & 73 & 72 & 70 & 66 & 61 & 58 & 55  \\
$N^{CC}_{sh} (10^\circ)$ & 66 & 112 & 124 & 122 & 119 & 112 & 104 & 99 & 94  \\
$N^{em}_{sh} (5^\circ)$ & 20 & 34 & 38 & 37 & 35 & 33 & 31 & 29 & 27  \\
$N^{em}_{sh} (10^\circ)$ & 34 & 58 & 65 & 63 & 60 & 56 & 53 & 49 & 46  \\
\hline
\hline
ATM$^{\mu}_{ct}$  &\multicolumn{4}{c}{839 ($5^\circ$)}&\multicolumn{5}{c}{3356 ($10^\circ$)} \\
ATM$^{\mu}_{up}$  &\multicolumn{4}{c}{564 ($5^\circ$)}&\multicolumn{5}{c}{2256 ($10^\circ$)} \\
ATM$^{NC}_{sh}$  &\multicolumn{4}{c}{169 ($5^\circ$)}&\multicolumn{5}{c}{676 ($10^\circ$)} \\
ATM$^{CC}_{sh}$  &\multicolumn{4}{c}{523 ($5^\circ$)}&\multicolumn{5}{c}{2092 ($10^\circ$)} \\
ATM$^{em}_{sh}$  &\multicolumn{4}{c}{34 ($5^\circ$)}&\multicolumn{5}{c}{136 ($10^\circ$)} \\
\hline
\end{tabular}
\caption{Event rates per km$^{2}$ per yr for the contained (ct), upward (u) muons ($\mu$)
and for the showers (sh) produced via charged current (CC), neutral
current (NC) and
electromagnetic
(em) interactions. Neutrinos from direct production ($\chi\chi\rightarrow\nu\overline\nu$) channel
and secondary neutrinos from $\chi\chi\rightarrow\tau^+\tau^-$ channel
are considered. We have set $B\cdot B_F =200$ for each channel. The cone half angle
is chosen to be 5$^\circ$ and 10$^\circ$. The threshold energy for the muon (shower)
events is set to be 80 (100) GeV. The backgrounds due to atmospheric neutrinos are also presented.}
\label{table:summary}
\end{table}

\section{Conclusion}

We have studied neutrino signals from dark matter annihilation in 
the Galactic Center.  We have calculated contained and upward 
muon fluxes from neutrino interactions, when neutrinos are 
produced in annihilation of dark matter either directly or 
via the decay of taus, W-bosons or b-quarks.  We have 
shown that in the case of direct neutrino production, the 
signal is above the atmospheric background for both contained 
and upward events, assuming that the annihilation rate 
is enhanced by boost factor of 200 (when the NFW dark matter 
halo profile is used) and that the branching 
ratio of dark mater annihilation into neutrinos is one. 
In general, the boost factor values that are required to explain the data 
obtained by the indirect detection experiments vary depending on the dark matter model and the dark matter mass. 
For the specific dark matter model 
our results can be rescaled by the 
corresponding product of the boost factor $B$ and the 
branching ratio $B_F$.

We have found that the contained muon flux dominates over the
upward muon flux for all energies when $m_\chi=200$ GeV.  
However, as we increase the mass $m_\chi$ of the dark matter particle, 
for example when $m_\chi = 500 $ GeV, the upward muon flux 
dominates up to $E_\mu = 300$ GeV, and for 
 $m_\chi = 800 $ GeV, up to 
 $E_\mu = 500$ GeV.  This is due to the increasing muon 
range as the muon initial energy increases, which becomes 
possible when $m_\chi$ is larger thus producing 
higher energy neutrinos in the annihilation.  
In the case of secondary neutrino production, the signal 
 becomes comparable to the background if the boost 
factor is an order of magnitude larger than the value 
we considered.  
We have shown that the shape of the muon flux depends on 
the specific decay mode, and that the dominant flux comes 
 from tau decay at low muon energies, and from W-decay for 
muon energies above 200 GeV.  
The total upward muon rates have a weak dependence on $m_\chi$ and 
on the muon energy threshold for 
$m_\chi > 400$ GeV, due to the balance of the energy dependence of the
muon range, the upper limit of the muon energy (given by $m_\chi$)  
 and the explicit dependence on $m_\chi$ ($ \sim m_\chi^{-2}$) 
of the muon  flux. However, the total contained muon rates show a sharp decrease with $m_\chi$
for $m_\chi>150$ GeV due to the finite size of the detector.  Upward 
muon events dominate over contained muon events for 
$m_\chi > 550$ GeV.

We have also shown that showers produced by neutrino interactions, 
when neutrinos are produced directly in dark matter annihilation, 
could also be used to detect a dark matter signal from the 
Galactic Center.  In particular, electromagnetic showers have 
much smaller background, from atmospheric electron neutrinos, 
than the hadronic showers. In addition, we have studied the contour plots of both the
upward muon events and the showers and we
have shown the required dependence of the annihilation cross section
on the dark matter mass in order to observe a fixed number of event rates. We have discussed the origin of 
 different shapes for the contour curves 
in each case and pointed out the contained event nature of the
shower events. We have shown that after one year 
 IceCube+DeepCore detector 
could potentially observe a 5$\sigma$ signal effect by measuring 
 contained muons (for direct neutrino production), or in 
$5$ to $8$ years a 2$\sigma$ effect with 
hadronic showers even in the case when 
they are due to secondary neutrinos.  

IceCube+DeepCore will be able to identify track-like events
due to the charged current interactions of muon neutrinos,  the showers due to
neutral current interactions of all the neutrino flavors and the charged current interactions of
electron and tau neutrinos. In particular, above the neutrino energy of 40 GeV the signal to background ratio for showers
is further enhanced since the atmospheric tau and electron neutrino fluxes are suppressed relative to the atmospheric 
muon neutrino flux.
Thus, the main background is the neutral current interaction
whose cross section is about a factor of three less than the charged current cross section
of the atmospheric muon neutrinos. 
The measurement of the ratio of track-like muon and shower events 
eliminates the dependence on some parameters of the theory (e.g., boost factor, the dark matter 
density
profile, etc) which only determine the overall normalization for the energy dependent differential muon fluxes, so 
the physical properties of the dark matter particle can better be determined. 

In addition to the boost factor due to Sommerfeld enhancement 
that we have considered, there is 
potential enhancement
of the dark matter signal due to the existence of small 
substructures in the Milky Way Halo \cite{savvas}.  
Possible observation of this 
additional 
boost may be difficult to observe because of the small population of 
these substructures
unless the neutrino detectors have a very good angular resolution 
 \cite{boost}.       

Due to its location in the northern hemisphere, the future KM3NeT experiment will be complementary to 
IceCube+DeepCore in searching for neutrino signals from dark matter annihilation in the Galactic Center through the observation of 
upward muon events. The atmospheric muon background  at the KM3NeT will be suppressed 
significantly since the Earth will act as a shield to those muons. Independent searches of the upward muon events by 
KM3NeT and the contained muon and shower events by IceCube+DeepCore look 
promising for the discovery of the mysterious 
dark matter particle or for setting stringent constraints on its properties.

\begin{acknowledgments}
We would like to thank Tyce DeYoung, Sven Lafebre, Irina Mocioiu, 
Anna Stasto and Tolga Guver for useful discussions.  IS and GG  would like to thank the Aspen
Center for Physics, where part of this work took place.
This research was supported by
US Department of Energy
contracts DE-FG02-91ER40664, DE-FG02-04ER41319 and DE-FG02-04ER41298.  
 GG was supported in part by the US Department of Energy Grant
DE-FG03-91ER40662, Task C at UCLA.
\end{acknowledgments}

\appendix
\section{Neutrino Energy Distributions}

\subsection{Neutrino energy distribution from direct production}

The neutrino energy distribution when neutrinos are produced directly from
dark matter annihilation is given by a delta function,

\begin{equation}
\frac{dN_\nu}{dE_\nu}=\delta(E_\nu-m_\chi)
\end{equation}  
where the assumption is that the dark matter particles are essentially at rest when they annihilate. 

\subsection{Neutrino energy distribution from $\tau^+\tau^-$ and 
$b\overline{b}$ decay
modes}
   
In these decay modes, we use the
unpolarized decay distributions, so the
$\nu$ and $\overline\nu$ distributions are assumed to be  
the same.
The decay
branching fraction is denoted by $B_f$ for a given decay mode $f$,
$f=\tau,b$. The $b$ quarks hadronize before they decay into
neutrinos. The hadronization effect is taken into account by scaling
the initial quark energy, $E_{in}=m_\chi$, in the form $E_{f}=z_f m_\chi $,
where $z_f=0.73$ for $b$ quarks\cite{jungman2}.  

The neutrino energy distribution from the decay of 
$f=\tau^+$, $\tau^-$, $b$ or $\overline{b}$ 
from $\chi\chi\to f\bar{f}$ is approximately
\begin{equation}\label{neutrino_dis}
\frac{dN_\nu}{dE_\nu}=\frac{2B_f}{E_{f}}(1-3x^2+2x^3),\;\;\; \mbox{where}\;\;\;
x=\frac{E_\nu}{E_{f}}\le 1\ ,
\end{equation}
where  for each neutrino or antineutrino flavor ($\nu_e$, $\bar{\nu}_e$,
$\nu_\mu$, $\bar{\nu}_\mu$), 
   \begin{equation}
(E_{f}\;,\;B_f)=\left\{
   \begin{array}{lr}
   (m_\chi\;,\;0.18) &           \tau \;\;\mbox{decay}, \\
   (0.73m_\chi\;,\;0.103) &       b\;\;  \mbox{decay}\ . 
   \end{array}
   \right.
\end{equation}

The energy distribution of the tau neutrinos from the decay of $f=b$ or $\overline{b}$
is given by (\ref{neutrino_dis}) and the distribution 
from the decay of $\tau^+$ or $\tau^-$ is given by \cite{Dutta},

\begin{equation}
\frac{dN_{\nu_\tau}}{dE_{\nu_\tau}}=\frac{4B_f}{3E_{f}}(1-x^3),\;\;\; \mbox{where}\;\;\;
x=\frac{E_{\nu_\tau}}{E_{f}}\le 1\ .
\end{equation}

\subsection{$W^+W^-$ decay mode}

In the $W^+W^-$ mode, when the dark matter particle is at rest when it
annihilates, $E_W = m_\chi/2$ and $\beta_W= \sqrt{1-m_W^2/m_\chi^2}$.
The decay distribution, for each $W$, is
\begin{equation}
\frac{dN_\nu}{dE_\nu}=\frac{B}{m_\chi\beta_W}\;\;\; \mbox{with} \;\;\;
\frac{m_\chi}{2}(1-\beta_W)<E_\nu<\frac{m_\chi}{2}(1+\beta_W)\ .
\end{equation}
\noindent
Here, $B=0.105$ for each neutrino flavor.

\section{Muon energy distribution}

The differential muon flux for the $\chi\chi\rightarrow\nu\overline\nu$ channel can 
be given as
\begin{eqnarray}
\nonumber
\frac{d\phi_\mu}{dE_\mu} =& & 
\frac{c}{m_\chi^2(E_\mu+\alpha/\beta)}\biggl[
a(m_\chi - E_\mu) \\
&+&\frac{b}{3 m_\chi^2}(m_\chi^3-E_\mu^3)\biggr]
\end{eqnarray}
where 
\begin{equation}
\label{functions}
c = B\frac{R_o \rho^2_o B_F \langle\sigma v\rangle_F
\langle J_2\rangle_\Omega\Delta\Omega
 m_{p}G^2_F N_A}{4\pi^2\beta}
\end{equation}
There is a separate distribution for neutrino and antineutrinos, since
the parameters $a$ and $b$ depend on the incident particle and the target.
Here, for isoscalar nucleon targets, $a=a_{\nu,\overline\nu}=0.20,0.05$ and $b=b_{\nu,\overline\nu}=0.05,0.20$. Also appearing are the Fermi
constant $G_F\simeq1.17\times10^{-5}$ GeV$^{-2}$ and Avogadro's number 
$N_A\simeq6\times10^{23}$. 
For standard rock, $\alpha \simeq 2\times 10^{-3}$ 
GeV\,cm$^2$/g accounts for the 
ionization energy loss and $\beta \simeq 3.0\times 10^{-6}$ cm$^2$/g accounts for the
bremsstrahlung, pair production and photonuclear interactions and we take $\rho=2.6$ g/cm$^3$. 

For the contained events, a similar expression can be derived as 
\begin{equation}
\frac{d\phi_\mu}{dE_\mu} = 
\frac{c'}{m^2_\chi}\left(a+b\frac{E^2_\mu}{m^2_\chi}\right)
\Theta(m_\chi-E_\mu)
\end{equation}
where $\Theta(x)=1$ if $x\ge0$ and $\Theta(x)=0$ otherwise, and 
\begin{equation}
c' = DB\frac{R_o \rho^2_o B_F \langle\sigma v\rangle
_F\langle J_2\rangle _\Omega\Delta\Omega m_p G^2_F N_A\rho}{4\pi^2}
\end{equation}
where $D$ is the size of the detector.

We note that 
\begin{eqnarray*}
\frac{d\phi_\mu}{dE_\mu} &\propto& \rho^0 \;\;\;\; \mbox{for the upward events} \\
\frac{d\phi_\mu}{dE_\mu} &\propto& \rho^1 \;\;\;\; \mbox{for the contained events} ,
\end{eqnarray*}
so, the muon flux doesn't depend on the rock density for the upward events except through $\alpha$ and $\beta$, whereas for the contained events, the muon flux is directly proportional to the density of 
the medium.

All the expressions for the muon flux derived below contain a  
$\Theta(m_\chi-E_\mu)$ function.
For secondary neutrinos which possess an energy spectrum in the form 
\begin{equation}
\left(\frac{dN}{dE}\right)_\nu = A\left(\frac{E_\nu}{m_\chi}\right)^n
\end{equation}
where A is an overall factor, the differential upward muon flux can be calculated by using

\begin{eqnarray}
\frac{d\phi_\mu}{dE_\mu} &=& 
\frac{cA}{m^{(n+2)}_\chi(E_\mu+\frac{\alpha}{\beta})} [P(m_\chi,E_\mu,n)+K(m_\chi,E_\mu,n)+\nonumber\\
&+&L(m_\chi,E_\mu,n)+M(m_\chi,E_\mu,n)]
\end{eqnarray}
where 

\begin{eqnarray}
P(m_\chi,E_\mu,n) &=& \frac{am^{(n+1)}_\chi(m_\chi-E_\mu)}{(n+1)}\nonumber\\
K(m_\chi,E_\mu,n) &=& -\frac{a(m^{(n+2)}_\chi-E^{(n+2)}_\mu)}{(n+1)(n+2)}\nonumber\\
L(m_\chi,E_\mu,n) &=& \frac{bm^{(n-1)}_\chi(m^3_\chi-E^3_\mu)}{3(n-1)}\nonumber\\
M(m_\chi,E_\mu,n) &=& -\frac{b(m^{(n+2)}_\chi-E^{(n+2)}_\mu)}{(n-1)(n+2)}. 
\end{eqnarray}
for $n\neq1$ and when $n=1$,

\begin{eqnarray}
\frac{d\phi_\mu}{dE_\mu} &=& \frac{cA}{3m^3_\chi(E_\mu+\frac{\alpha}{\beta})}\times\nonumber\\
&\times&[m^3_\chi\left(a+\frac{b}{3}\right)-\frac{3aE_\mu m^2_\chi}{2}+\nonumber\\
&+& E^3_\mu\left(b\ln\left(\frac{E_\mu}{m_\chi}\right)+\frac{a}{2}-\frac{b}{3}\right)].
\end{eqnarray}

For the contained events and when $n\neq1$,

\begin{eqnarray}
\frac{d\phi_\mu}{dE_\mu} &=& \frac{c'A}{m^{(n+2)}_\chi}[\frac{a}{(n+1)}(m^{(n+1)}_\chi-E^{(n+1)}_\mu)+\nonumber\\
&+& \frac{bE^2_\mu}{(n-1)}(m^{(n-1)}_\chi-E^{(n-1)}_\mu)]
\end{eqnarray}
which reduces to 

\begin{eqnarray}
\frac{d\phi_\mu}{dE_\mu} &=& \frac{c'A}{m^{3}_\chi}[\frac{a}{2}(m^{2}_\chi-E^{2}_\mu)+\nonumber\\
&+& bE^2_\mu \ln\left(\frac{m_\chi}{E_\mu}\right)].
\end{eqnarray}
when $n=1$.

\end{document}